\documentstyle[12pt,aasms4]{article}

\received{}
\accepted{}

\lefthead{Ribas {\em et al.}}
\righthead{The LMC eclipsing binary HV 2274}

\begin{document}

\title{The LMC eclipsing binary HV 2274: fundamental properties \\
       and comparison with evolutionary models}

\author{Ignasi Ribas}
\affil{Departament d'Astronomia i Meteorologia, Universitat de Barcelona,\\
Av. Diagonal, 647, E-08028 Barcelona, Spain\\
E-mail: iribas@am.ub.es}

\author{Edward F. Guinan, Edward L. Fitzpatrick, Laurence E. DeWarf,\\
Frank P. Maloney, Philip A. Maurone}
\affil{Department of Astronomy and Astrophysics, Villanova University,\\
Villanova, PA 19085, USA\\
E-mail: guinan,dewarf,maurone@ucis.vill.ed; fitz@ast.vill.edu; 
fmaloney@email.vill.edu}

\author{David H. Bradstreet}
\affil{Department of Physical Science, Eastern College, St. Davids, PA 19087,
USA\\
E-mail: bradstre@beacon.eastern.edu}

\author{\'Alvaro Gim\'enez}
\affil{LAEFF (INTA-CSIC), Apartado 50727, E-28080 Madrid, Spain \\
E-mail: ag@laeff.esa.es}

\and

\author{John D. Pritchard}
\affil{Mount John University Observatory and Department of Physics \& 
Astronomy, \\
University of Canterbury, Private Bag 4800, Christchurch, New Zealand\\
E-mail: j.pritchard@phys.canterbury.ac.nz}

\begin{abstract}

We are carrying out an international, multi-wavelength program to
determine the fundamental properties and independent distance estimates
of selected 14$^{\rm th}$ to $15^{\rm th}$ magnitude eclipsing binaries
in the Large and Small Magellanic Clouds (LMC and SMC).  Eclipsing
binaries with well-defined double-line radial velocity curves and light
curves provide valuable information on orbital and physical properties
of their component stars.  These properties include, among other
characterisitcs, stellar mass and radius.  These can be measured with
an accuracy and directness unachievable by any other means. The study
of stars in the LMC and SMC where the metal abundances are
significantly lower than solar (by 1/3 to 1/10) provides an important
opportunity to test stellar atmosphere, interior and evolution models,
and opacities.  For the first time, we can also measure direct
Mass-Luminosity relations for stars outside our Galaxy.
 
In a previous paper we demonstrated how a precise distance to the LMC --- 
corresponding to $(V_{\circ}-M_v)_{\rm LMC}=18.30\pm0.07$ mag --- could be 
determined using the 14$^{\rm th}$ mag LMC eclipsing binary HV 2274.  In 
this paper we concentrate on the determination of the orbital and physical 
properties of HV 2274 and its component stars from analyses of light curves 
and new radial velocity curves formed from HST/Goddard High-Resolution 
Spectrograph observations. HV 2274 (B1-2 IV-III + B1-2 IV-III; $V_{\rm max} 
\approx +14.2$; $P = 5\fd73$) is a particularly appealing star because it is 
a detached binary which has an eccentric orbit ($e=0.136$) and shows fast
apsidal motion. The results of these analyses yield reliable masses and
absolute radii, as well as other physical and orbital properties of the
stars and the system.  From UV/optical spectrophotometry
($1150-4820$~\AA) of HV 2274 obtained with the HST/Faint Object
Spectrograph, the temperatures and the metallicity ($[Fe/H]=-0.45\pm
0.06$) of the stars were found, as well as the interstellar extinction
of the system.  The values of mass, absolute radius, and effective
temperature, for the primary and secondary stars are:
$12.2\pm0.7$~M$_{\odot}$, $9.9\pm0.2$~R$_{\odot}$, $23000\pm180$~K, and
$11.4\pm0.7$~M$_{\odot}$, $9.0\pm0.2$~R$_{\odot}$, $23110\pm180$~K,
respectively.

The age of the system ($\tau = 17\pm2$~Myr), helium abundance
($Y=0.26\pm0.03$) and a lower limit of the convective core overshooting
parameter of $\alpha_{\rm ov} \approx 0.2$ were obtained from fitting
the stellar data with evolution models of Claret \& Gim\'enez. HV 2274
has a relatively well-determined (and fast) apsidal motion period of
$U=123\pm3$~yr.  From the analysis of apsidal motion, additional information 
and constraints on the structure of the stars can be made. The apsidal motion 
analysis corroborates that some amount of convective core overshooting 
($\alpha_{\rm ov}$ between 0.2 and 0.5) is needed.

\end{abstract}


\section{Introduction}

Extragalactic eclipsing binaries, and in particular systems belonging
to the Large and Small Magellanic Clouds (LMC and SMC), can be used to
probe the structure and evolution of stars in environments with
chemical histories that differ significantly from those of the solar
neighborhood.  Studying massive, metal deficient stars in the LMC and
SMC is like using a ``time machine'' for studying the low-metallicity
massive old disk O- and B-type stars that once populated our Galaxy about
5-10 billion years ago. In addition, much needed checks of the
evolutionary models for these extreme metal abundances, which are
widely used in stellar population synthesis calculations, can be
performed.  LMC and SMC eclipsing binaries can also be used to address
other astrophysically --- and cosmologically --- important issues, such
as the structure and dynamics of these galaxies, the enrichment of the
interstellar medium, and the distance of the LMC, an important rung on
the Cosmic Distance Ladder.  General discussions of the importance of
extragalactic eclipsing binaries can be found in Graham (1983), Davidge
(1987), Koch (1990) and Guinan (1993).
 
The LMC eclipsing binary HV~2274 ($V_{\rm max} \approx +14.2$) is one
of the most interesting and important systems in our current program of
LMC and SMC eclipsing binaries (see Guinan et al. 1996, 1998a). HV~2274
is an uncomplicated, detached system consisting of two nearly-identical
B1-2 IV-III stars, moving in an eccentric orbit with a period of
5\fd73. Moreover, it lies in an uncrowded field, so there is no
contamination of the light curves by neighboring stars. $BVI$ light curves
have been obtained by Watson et al. (1992, hereafter W92). They modeled
the light curves and have determined preliminary values of the
photometric, orbital, and geometric properties of the system.
 
One of the most interesting aspects of HV~2274 is its eccentric orbit,
which allows the apsidal motion of the system to be measured.  W92
combined recent CCD eclipse timings with older eclipse times estimated
from the early photographic photometry and, from this baseline of
nearly 60 years, measured an apsidal motion period of U=123$\pm$3~yr.
Eclipsing binaries with well-determined apsidal motion rates are
important because the internal mass distributions of the stars can be
measured (e.g., see Kopal 1959; Guinan 1993). These direct measurements
provide extremely important checks on stellar interior and evolution
models.

Recently, Guinan et al. (1998b) showed how the combination of light and
velocity curve information with UV/optical spectophotometry could yield
a precise distance to the HV 2274 system, and thus to the LMC. In
this paper, we highlight the properties of the stars themselves and
discuss the constraints placed on stellar interiors models through
analysis of HV 2274.  In \S 2 the data used in this study are
described. \S 3 is devoted to the modelling of the observational data
that leads to the absolute dimensions of the system components. 
These fundamental properties are compared with the predictions of 
evolutionary models in \S 4. The tidal evolution of the system, i.e.
synchronization and circularization times, is studied in \S 5.
And, finally, \S 6 summarizes the main conclusions derived from this 
work.

\section{Observational data}

Johnson $BVI$ light curves of HV~2274 were obtained by W92. The observations 
were made in 1989--1990 from Mount John Observatory (New Zealand) with a 
0.61~m telescope equipped with a CCD. Differential photometry in all $BVI$ 
filters was obtained over 20 nights, with a total number of about 110 
measurements in each of the passbands. 

\begin{figure*}
\epsscale{1.00}
\plotone{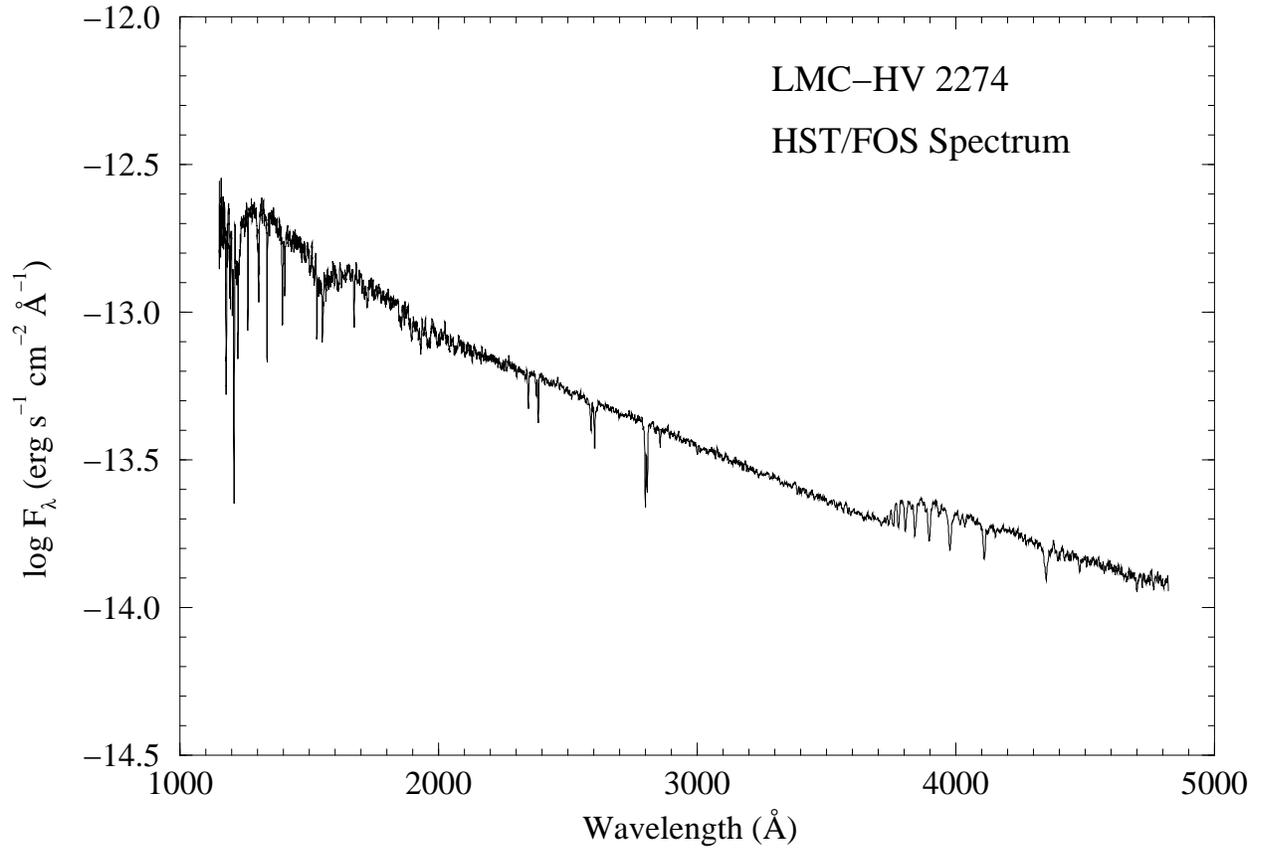}
\figcaption[fig1.eps]{Full observed HST/FOS spectrum of HV~2274 
(5-point smoothing filter applied).
\label{fig:figFOSspec}}
\end{figure*}

HV~2274 was observed with HST as part of a project to obtain UV
spectrophotometry of eclipsing binaries in the Magellanic Clouds.  Four
low-resolution spectra, covering four adjacent wavelength regions, were
acquired with HST/FOS (Faint Object Spectrograph).  The four spectra
were obtained during two successive 96 min. HST orbits, i.e. at
essentially the same orbital phase ($\pm0\fp02$).  It is important to
notice that the studied spectral region contains about 80\% of the
total flux of the stars.  The main characteristics of the spectra are
shown in Table \ref{tab:logFOS}. The raw data were automatically flux
and wavelength calibrated using a set of IDL-based routines prepared by
the STScI. The calibration accounted for all the instantaneous
information about the spacecraft status, diode problems and instrument
response. The data on different wavelength regions were merged (and
averaged in the overlapping zones) to produce a single spectrum
spanning 1150~\AA~to 4820~\AA~which is shown in Fig.
\ref{fig:figFOSspec}.

\placetable{tab:logFOS}
\placefigure{fig:figFOSspec}

HV~2274 was also selected as a target for a pilot study to secure
radial velocity observations with HST so that accurate masses and
absolute radii of the stars could be measured. Fourteen spectra were
taken at different orbital phases using HST/GHRS (Goddard
High-Resolution Spectrograph) in its medium-resolution mode (R=23000,
14~km~s$^{-1}$/diode, $\Delta \lambda = 34$~\AA), with integration
times $\sim$1100~s and a S/N ratio of about 20. The spectra, whose
characteristics are listed in Table \ref{tab:logGHRS}, cover a
wavelength range of 34~\AA~in two different spectral regions, centered
at about $\lambda$1305~\AA~and $\lambda$1335~\AA, where strong UV
stellar photospheric lines are typically present for early B-type
stars. This fact clearly demonstrates the advantage of UV space-based 
observations over optical ground-based observations. In the UV 
domain, numerous photospheric absorption features can be easily 
identified and used, e.g. \ion{Si}{3}, \ion{C}{2}, and ionized 
iron-peak elements. In contrast, in the optical range, even with echelle 
spectra 3000~\AA~wide, only a handful of moderately strong \ion{He}{1} 
lines are suitable for obtaining radial velocity measurements.

\begin{figure*}
\epsscale{1.00}
\plotone{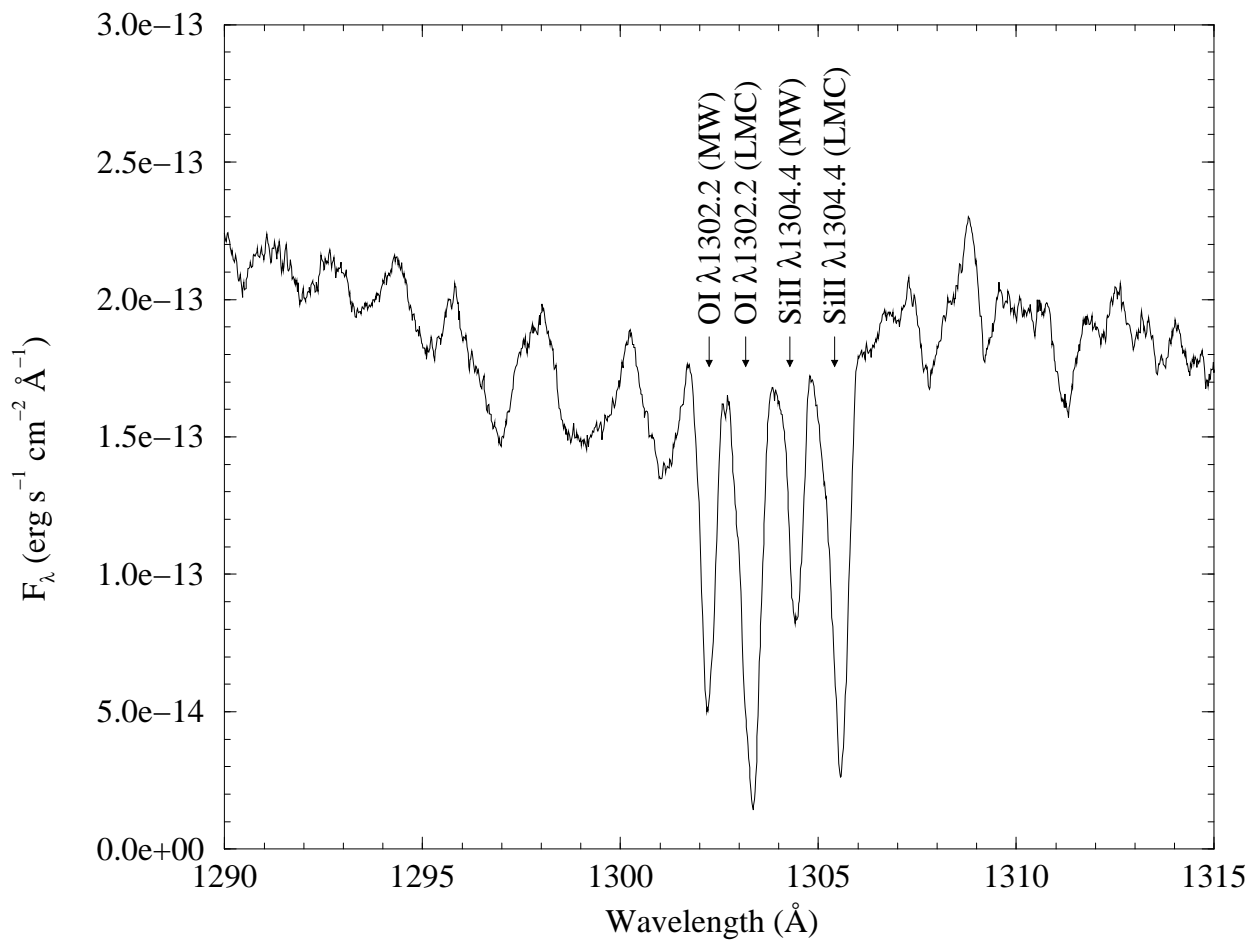}
\figcaption[fig2.eps]{HST/GHRS spectrum of HV~2274 centered at the 
$\lambda$1300~\AA~region (spectrum Nr. 12 in Table \ref{tab:logGHRS}). The 
strongest features in the spectrum are ISM absorption lines from both our 
Galaxy (MW) and the LMC. The lines are labeled with the ion identification 
and the rest wavelength.
\label{fig:figGHRStot}}
\end{figure*}

One of the complete HST/GHRS spectra is shown in Fig. \ref{fig:figGHRStot}  
as an example. The spectra taken at orbital conjunction (phases close to 
0\fp0 and 0\fp6) were not used for radial velocity determinations since the 
lines of the components could not be resolved. The spectra contain a number 
of interstellar lines that are useful for wavelength calibrations.

\placetable{tab:logGHRS}
\placefigure{fig:figGHRStot}

\begin{figure*}
\epsscale{1.00}
\plottwo{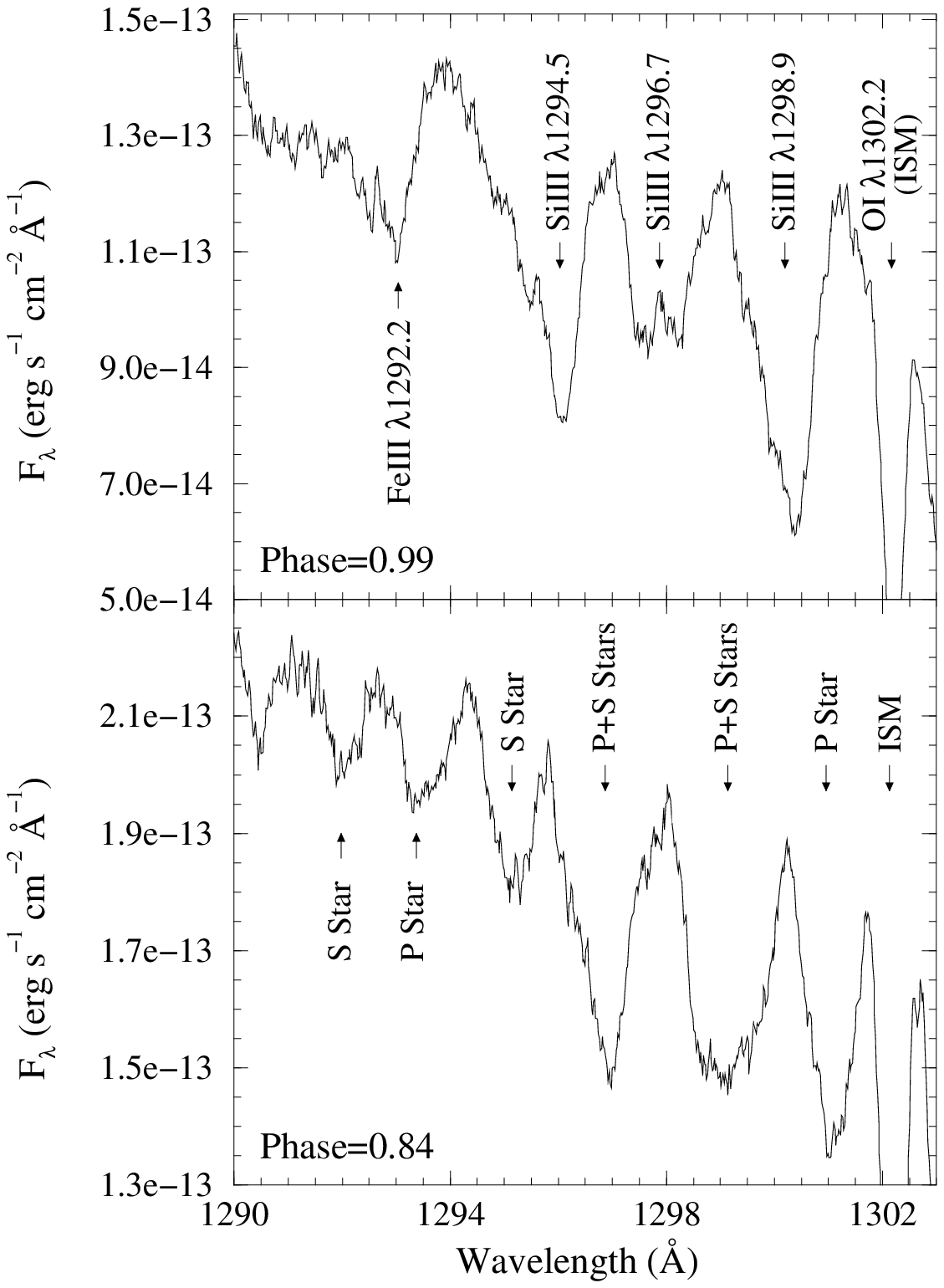}{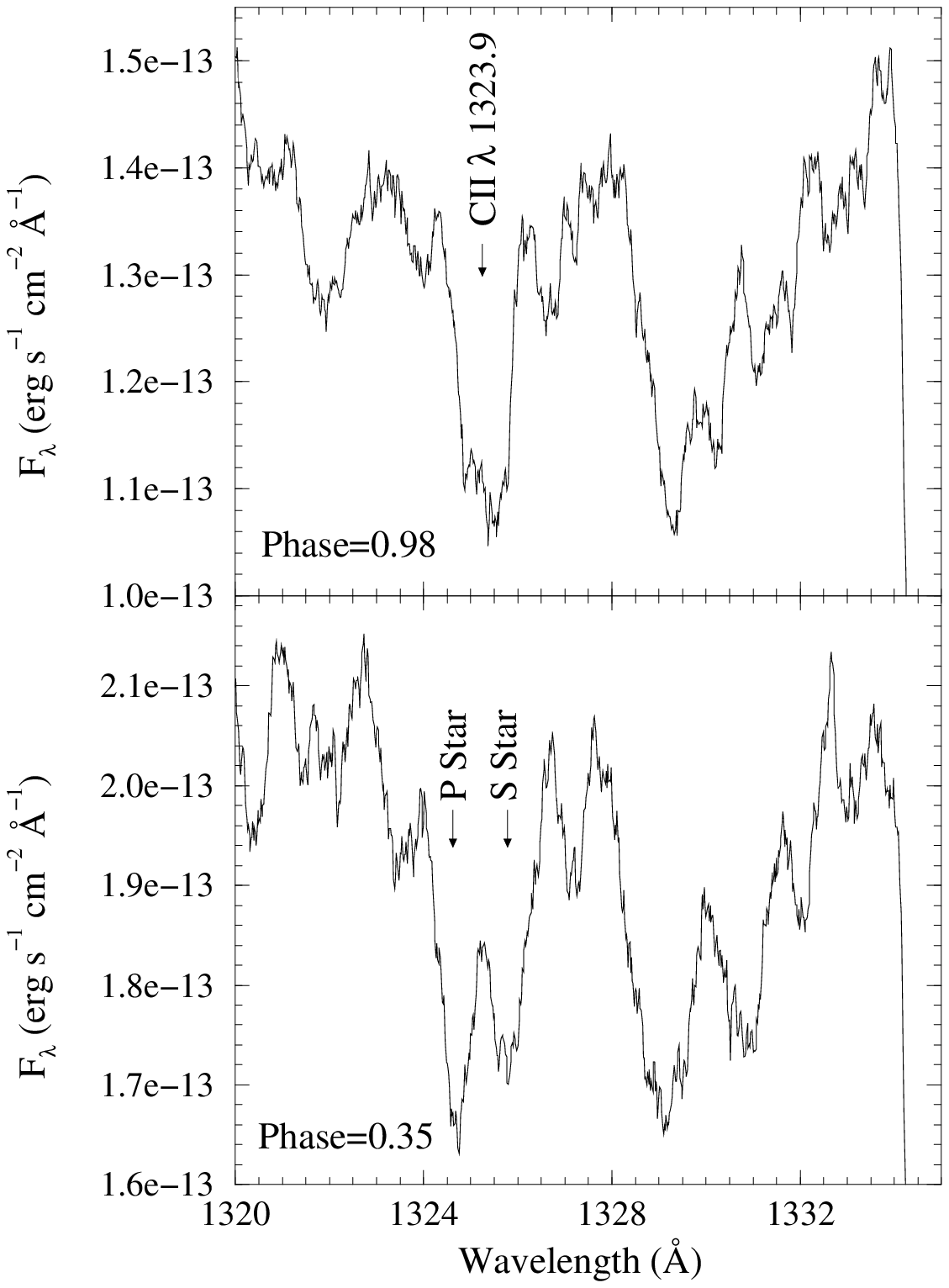}
\figcaption[fig3a.eps,fig3b.eps]{HST/GHRS spectra of HV~2274. The upper panels 
show two spectra taken at phases 0\fp99 and 0\fp98 (during primary eclipse, 
spectra Nr.  4 and 3 of Table \ref{tab:logGHRS}) where single lines of 
\ion{Fe}{3}, \ion{Si}{3} (triplet) and \ion{C}{2} are identified. The spectra 
shown in the lower panel were taken at phases 0\fp84 and 0\fp35 (spectra Nr. 11 
and 13 of Table \ref{tab:logGHRS}), and the lines corresponding to the primary
(P) and secondary (S) components are observed, though in some cases (\ion{Si}{3} 
triplet) they overlap. 
\label{fig:figGHRS2}}
\end{figure*}

Stellar lines of \ion{Fe}{3} at $\lambda$1292.2~\AA, \ion{Si}{3}
(triplet) at $\lambda$1294.5, $\lambda$1296.7 and $\lambda$1298.9~\AA,
and \ion{C}{2} at $\lambda$1323.9~\AA~were identified in the spectra,
as well as several strong ISM lines from both the Milky Way and the
LMC. In Fig. \ref{fig:figGHRS2} we present, as an example, the
comparison of two enlarged spectra taken at different phases, with the
line identifications included. Several tests were performed to obtain
the radial velocity of each component using cross-correlation
techniques, but both the small number of usable lines and the strong
ISM contamination prevented the obtaining of good results.  Instead,
double Gaussian fits were applied to derive the exact center of each
line for each component and the corresponding radial velocity shift.
The continuum level and the intensity, the center and the FWHM of the
lines were fitted to the observed spectrum. The fits were, in general,
very good. The measurements of different lines in a single spectrum
were then averaged to obtain the radial velocity at each phase. The
radial velocity measures for each star are listed in Table
\ref{tab:RVHV2274} and are plotted against orbital phase in Fig.
\ref{fig:figRVcurve}. The ephemeris of W92 was used to compute the
orbital phases. The mean uncertainties in the radial velocity measures
are approximately $\pm$15~km~s$^{-1}$.

\placefigure{fig:figGHRS2}
\placetable{tab:RVHV2274}

\section{Analysis and results}

All spectroscopic and photometric information was combined to obtain a 
reliable determination of the orbital, physical and radiative parameters of 
the system and its components. An iterative procedure was used to achieve 
internal consistency among the solutions of the light and radial velocity 
curves and the UV spectrophotometry fit.

\subsection{Modeling the light and radial velocity curves}

The radial velocity and light curves were both solved simultaneously using an 
improved Wilson-Devinney program (Wilson \& Devinney 1971; Wilson 1979, 1990) 
that includes the Kurucz (1979) atmosphere model routine developed by Milone 
et al. (1992, 1994). An automatic procedure was developed in order to keep the 
WD program running until a solution is found (see Pritchard et al. 1998 for a 
complete discussion). In this procedure, the computed differential 
corrections are applied to the input parameters in order to start a new 
iteration.

The free parameters in the curve fitting were: the eccentricity ($e$),
the longitude of the periastron ($\omega$), the phase offset
($\phi_{\circ}$), the inclination ($i$), the temperature of the
secondary ($T_{\rm S}$), the gravitational potentials ($\Omega_{\rm P}$
and $\Omega_{\rm S}$), the luminosity of the primary ($L_{\rm P}$), the
orbital semi-major axis ($a$), the mass ratio ($q=M_{\rm S}/M_{\rm
P}$), and the systemic radial velocity ($\gamma$). Several initial
tests done by solving for all orbital, physical and radiative
parameters at the same time (light and radial velocity simultaneously),
showed clear systematics in the residuals of the light curve because of
the small value of the eccentricity indicated by the sparsely covered
radial velocity curve. This may also be caused by the fast apsidal
motion rate and difference in the epoch of the photometric and radial
velocity observations. Due to the correlation, the eccentricity and the
longitude of the periastron were derived from the light curve solution
alone and held constant in all the subsequent trials.  The radial
velocity curve and the light curve were analyzed separately and several
iterations were performed until reaching a mutually consistent
solution.

\begin{figure*}
\epsscale{1.00}
\plotone{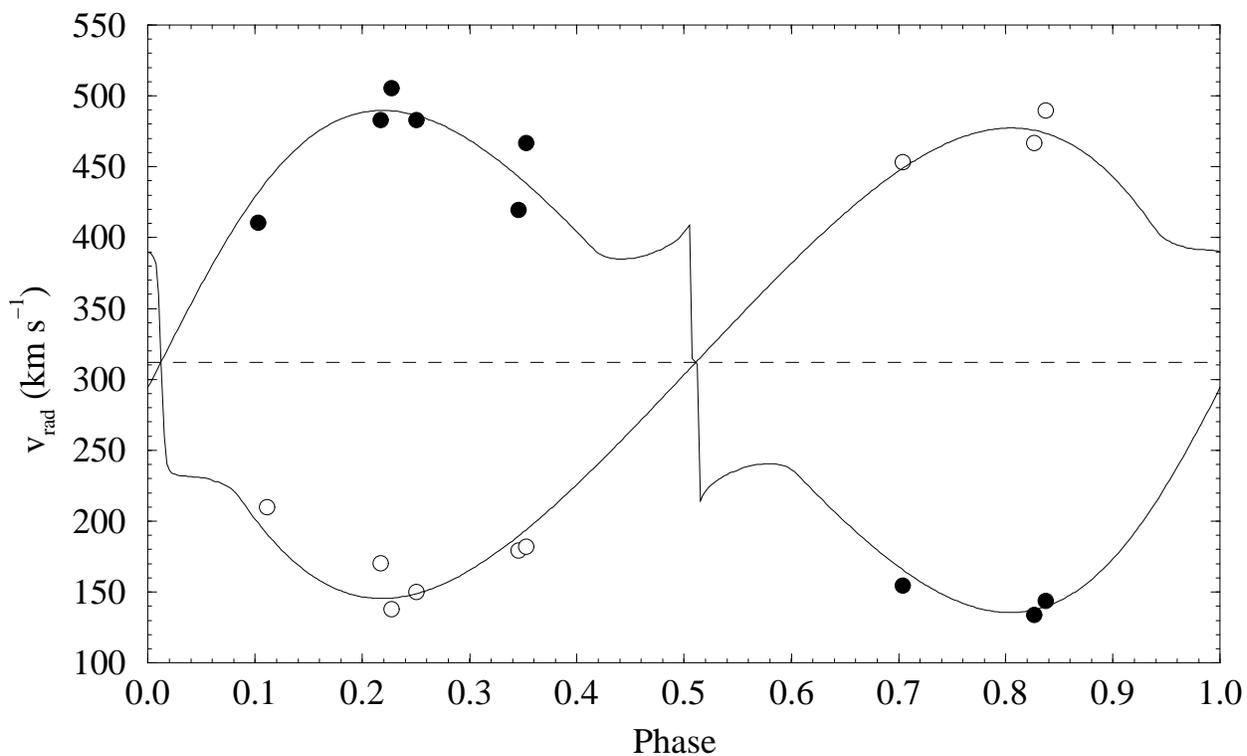}
\figcaption[fig4.eps]{Radial velocity observations of HV~2274 from 
medium-resolution HST/GHRS spectra (Table \ref{tab:RVHV2274}). The empty 
circles correspond to the primary component whereas the measurements of the 
secondary component are represented as filled circles. The best fitting 
curve was obtained by modeling with WD program. The parameters $e$ and $\omega$ 
were fixed to the values obtained from the light curve solution. The jumps seen 
in the model curves, known as ``Rossiter effect'', occur when rotating stars 
are partially eclipsed. 
\label{fig:figRVcurve}}
\end{figure*}

The fitting of a radial velocity curve is conceptually a simple problem. The 
free parameters were the orbital semi-major axis ($a$), the mass ratio ($q$) 
and the systemic velocity ($\gamma$), whereas the eccentricity ($e$) and 
longitude of the periastron ($\omega$) were set constant to the value given by 
the light curve solution. We shall remark that $\omega$ was corrected for the 
epoch difference by using the apsidal motion rate determination of W92. The 
best fitting curve is shown in Figure \ref{fig:figRVcurve} and Table
\ref{tab:RVHV2274} lists the individual velocity residuals, which have an
rms scatter of 14~km~s$^{-1}$ and 17~km~s$^{-1}$  for the primary and
secondary components, respectively. The resulting values of the radial 
velocity curve analysis are included in Table \ref{tab:HV2274}. The systemic 
radial velocity that we obtain, $\gamma=+312$~km~s$^{-1}$, is in good 
agreement with the currently adopted value for the heliocentric radial 
velocity of the LMC galaxy of $+300$~km~s$^{-1}$. Moreover, the HST/GHRS 
spectra show strong ISM lines from LMC gas, identified as \ion{O}{1} at 
$\lambda$1302.2~\AA~and \ion{Si}{2} at $\lambda$1304.4~\AA, that show only 
one component with a heliocentric radial velocity of $+285$~km~s$^{-1}$. The 
heliocentric radial velocity of the foreground galactic ISM lines was 
measured to be $-6$~km~s$^{-1}$.

\placefigure{fig:figRVcurve}

A detached configuration for the binary (i.e. stars inside their Roche lobes),
with coupling between luminosity and temperature was chosen when running the 
light curve solutions, as suggested by the uncomplicated shapes of the light 
curve outside the eclipses. Reflection and proximity effects were considered 
although they are expected to be unimportant for this well-detached system. 
Both the bolometric albedos and gravity darkening exponents were set to 1.0 
as currently adopted for the radiative atmospheres of the components (see 
Wilson et al. 1972). The mass ratio $q$ was fixed to the spectroscopic value 
and the temperature of the primary star was set to 23000~K as deduced from the 
UV/optical spectrophotometry fit (see next section). Finally, the rotational
velocities of the components were adopted as $125$~km~s$^{-1}$ and 
$115$~km~s$^{-1}$ (see \S 3.3).

The $BVI$ light curves were fit simultaneously in order to achieve a 
single, mutually consistent solution. At each iteration of the automatic 
procedure, the differential corrections were applied to the input parameters 
to build the new set of parameters for the next iteration. The result was 
carefully checked in order to avoid possible unphysical situations (e.g. 
Roche lobe filling in detached configuration). A solution was defined as the 
set of parameters for which the differential corrections suggested by the WD 
program were smaller than the standard errors during three consecutive 
iterations. However, when a solution was found, the program did not terminate. 
Instead, it was kept running in order to test the stability of the solutions, 
to evaluate their scatter, and to check for possible spurious solutions. 

In the modeling of the light curves of eccentric systems, there exist two 
different sets of parameters that provide almost identical fits to the 
observations. For HV~2274, the first one has a more massive star which is 
larger and cooler than the less massive star, whereas in the other solution, 
the more massive component is smaller and hotter. There is not sufficient 
information in the light or radial velocity curves to discriminate between 
these two possible solutions, and other data sources have to be considered to 
break this degeneracy. Since the luminosity ratio between the components 
predicted by the two solutions is very different ($L_{\rm S}/L_{\rm P} 
\approx0.85$ and $L_{\rm S}/L_{\rm P}\approx1.2$), we measured the equivalent 
widths of several pairs of lines in the GHRS spectra to estimate the 
spectroscopic light ratio of the two stars. We obtained a mean value of 
$0.88\pm0.05$, which, given the nearly identical temperatures of the stars, 
clearly indicates that the first solution is the physically 
preferable one.  

\begin{figure*}
\epsscale{1.00}
\plotone{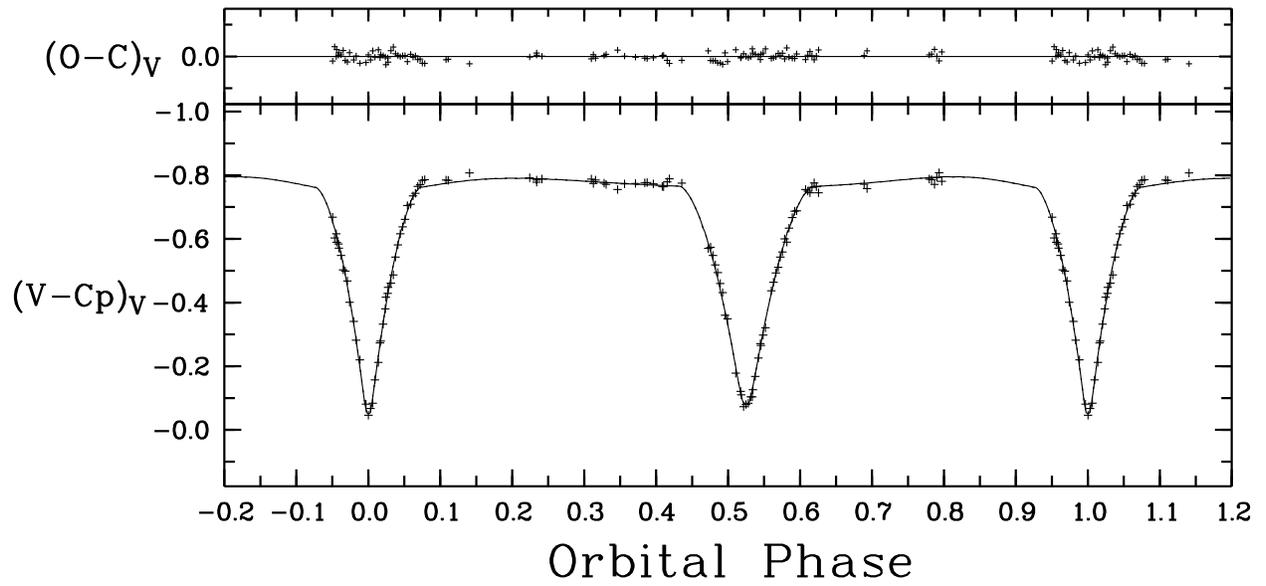}
\figcaption[fig5.eps]{Light curve fit to the observed $V$ Variable$-$Comparison 
(V$-$Cp) differential photometry of W92. Also shown are the 
Observed$-$Computed (O$-$C) residuals. The orbital phase is computed 
according to the linear ephemeris $\mbox{Min I (HJD)} = 2448099.818 + 5.726006 
\, E$.
\label{fig:figLC}}
\end{figure*}

The best fitting parameters are shown in Table \ref{tab:HV2274}. With
such parameters, the rms scatters of the residuals are 0.018 (n=108),
0.013 (n=116) and 0.019~mag (n=112) for $BVI$ respectively. Fig.
\ref{fig:figLC} shows, as an example, the light curve fit to the
observed $V$ differential photometry and the corresponding residuals.
The uncertainties in the light curve parameters were carefully
evaluated and adopted as three times the rms scatter of a set of
several solutions obtained from different initial conditions, instead
of the formal standard errors computed by the WD program.  This
approach provides more realistic estimations of the actual
uncertainties of the computed properties.

\placefigure{fig:figLC}

W92 also attempted a preliminary solution of their observed light
curves by using the EBOP program (Etzel 1975, 1993). Since
they had no information about the mass ratio ($q$) or the radius ratio
($k$), only the orbital parameters $e$, $i$ and $\omega$, that are
essentially insensitive to $q$ and $k$, were explicitly listed by the
authors. When comparing our results with their initial estimations we
find agreement within the quoted errors.

\subsection{Modeling the spectrophotometry}

The spectrophotometric analysis consisted of finding the pair of ATLAS9
atmosphere models (Kurucz 1991, 1994) which best reproduces the
observed energy distribution of the HV~2274 system (FOS data plus
$B-V$). The $B$ and $V$ calibrated photometry outside the eclipses was
taken from the photometric measurements of Udalski et al. (1998). The
photometry of W92 was not used because of its poorer accuracy ($\approx
0.1$~mag), as pointed out by the authors themselves.

Each ATLAS9 model is characterized by four parameters: effective
temperature ($T_{\rm eff}$), surface gravity ($\log g$), metalicity
($[m/H]$) and microturbulence velocity ($\mu$).   A non-linear least
squares algorithm developed by Fitzpatrick \& Massa (1999) was employed
to find the combination of model parameters --- along with the
appropriate interstellar extinction curve --- which best reproduce the
observations.  The technique is discussed in detail in Fitzpatrick \&
Massa (1999), but, in short, this approach is feasible because the
signatures of interstellar extinction, temperature, surface gravity,
metallicity and microturbulence in the model spectrum are very
different from each other.  The fitting algorithm also evaluates the
1$\sigma$ internal errors that account for the full interdependence of
all the parameters.  

The eight possible parameters describing the HV2274 models are heavily
constrained by the results of the light and radial velocity curve
analysis, which yielded values of the surface gravity of each component
and the ratio of the effective temperatures.  As additional constraints
we assumed that the two stars in the system have identical metallicity
and microturbulence velocity.

\begin{figure*}
\epsscale{1.00}
\plotone{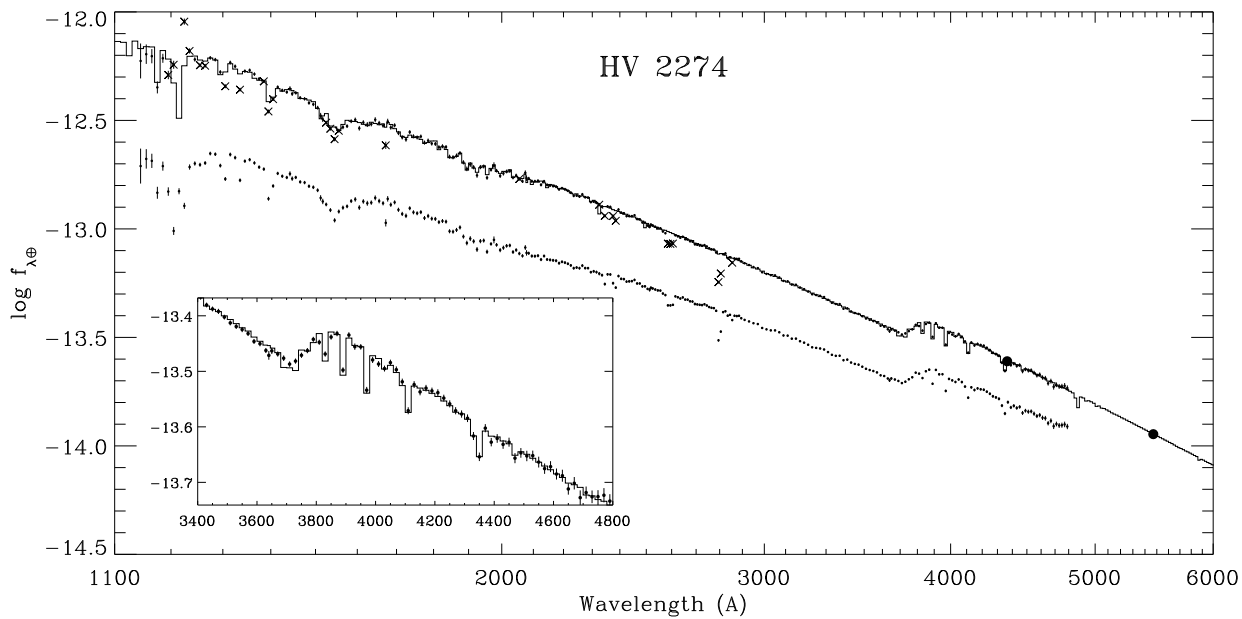}
\figcaption[fig6.eps]{The UV/optical energy distribution of HV~2274, in 
units of erg~cm$^{-2}$~s$^{-1}$~\AA$^{-1}$. The lower full spectrum shows the 
observed HST/FOS energy distribution; the upper spectrum shows the 
extinction-corrected energy distribution, superimposed with the best-fitting
ATLAS9 atmosphere model (plotted in histogram style). Vertical lines
through the data points indicate the 1$\sigma$ observational errors. 
Crosses indicate data points excluded from the fit due to contamination
by interstellar absorption lines. The large filled circles show the 
dereddened B and V photometry from Udalski et al. (1998). An expanded 
view of the fit near the Balmer jump is shown within the inset.
\label{fig:figFOSfit}}
\end{figure*}

The best fitting model to HV2274 is shown  in Fig. \ref{fig:figFOSfit}
and Table \ref{tab:HV2274} summarizes the derived properties, together
with their errors. The shape of the UV interstellar extinction curve,
derived via the fitting procedure, shows a much weaker
2175~\AA~extinction bump than found along typical Milky Way lines of
sight.  The properties of the UV extinction curves seen towards
Magellanic Cloud eclipsing binaries will be discussed in a future
paper.  To investigate the validity of the ATLAS9 models and, thus, the
physical significance of the derived quantities, we performed a second
fit by constraining only the difference in surface gravities between
the components but not the actual $\log g$ values.  This resulted in a
value of $\log g=3.51\pm0.05$~dex for the primary component,
gratifyingly close to the result based on the orbital solution alone.
In addition, the metal abundance from the FOS spectrum fit, $[Fe/H] =
-0.45\pm0.06$~dex with respect to solar abundances, is in good
agreement with the metal abundances typically reported for the LMC
($\sim$ $-0.5$, $-0.6$~dex; Bica et al. 1998). These results
strengthen our confidence in the determination of the remaining
parameters.

\placefigure{fig:figFOSfit}
\placetable{tab:HV2274}

\subsection{Rotational velocity}

Apart from obtaining a radial velocity curve, the HST/GHRS spectra were
used to estimate the rotational velocity of the components. An accurate 
measurement was not possible due to the lack of spectra of standard stars, 
but an estimation could be done by comparing the observations with 
synthetic spectra. When building the synthetic spectra, we adopted ATLAS9 
models and the set of programs developed by I. Hubeny ({\sc synspec}, {\sc 
rotin} and {\sc synplot}), which were modified to combine, in the correct 
proportion, the flux of both components (see Ribas et al. 1999). Atmosphere 
models with $\log~g=3.5$~dex were used. The effective temperature of each 
component, the microturbulence velocity and the metalicity were fixed from 
the results of the spectrophotometry fit. 

The rotational velocities of the components, the only free parameters, were 
interactively changed until the best possible agreement between the observed 
and the synthetic spectra was reached. The profiles of the same lines used
for radial velocity measures (mentioned in \S 2) were employed during this
procedure. The best fit to the observed spectrum was obtained when considering 
rotational velocities of $125$~km~s$^{-1}$ and $115$~km~s$^{-1}$ for the 
primary and secondary components, respectively. The error of these values is 
estimated to be around $10$~km~s$^{-1}$. The pseudo-synchronization 
velocities (i.e. synchronization at periastron) for the components of HV~2274 
are $117$~km~s$^{-1}$ and $107$~km~s$^{-1}$, both within the error bars of
our estimates. Therefore, we conclude that both components are synchronized
with the orbital velocity at periastron.

\section{Comparison with Evolutionary Models}

A comparison of the derived physical properties of the stars with the 
theoretical predictions of stellar evolutionary models was also made. We
considered the evolutionary models of Claret (1995, 1997) and Claret \& 
Gim\'enez (1995, 1998) (altogether referred to as CG). These models cover a 
wide range in both metalicity ($Z$) and initial helium abundance ($Y$), 
incorporate the most modern input physics and adopt a value of 0.2~H$_{\rm p}$ 
as the overshooting parameter ($\alpha_{\rm ov}$). An identical set of models 
that consider no convective overshoot was kindly provided by A. Claret (priv.
comm.).

The stellar evolutionary models were interpolated at a metal abundance
of $Z=0.007$, as suggested by the results of the FOS spectrum fit. Once
the metalicity is fixed and the masses, radii and effective
temperatures of the components are known, the only free parameter in
the the adopted set of evolutionary models is the initial helium
abundance $Y$. A simple least squares procedure indicated that the
value of the initial helium abundance that best reproduces the current
properties of HV~2274 components is $Y=0.26$. After several more trials
taking into account the uncertainties in masses, radii and effective
temperature, we estimated the uncertainty of this quantity to be around
$\pm0.03$. The main source of error in the determination of $Y$ is the
uncertainty in the masses, but nevertheless, the good accuracy of the
estimation, $Y=0.26\pm0.03$, is noteworthy. This value of $Y$ shows a
very good agreement with the expected helium abundance from the
currently adopted chemical enrichment law (see Izotov \& Thuan 1998).
 
\begin{figure*}
\epsscale{1.00}
\plotone{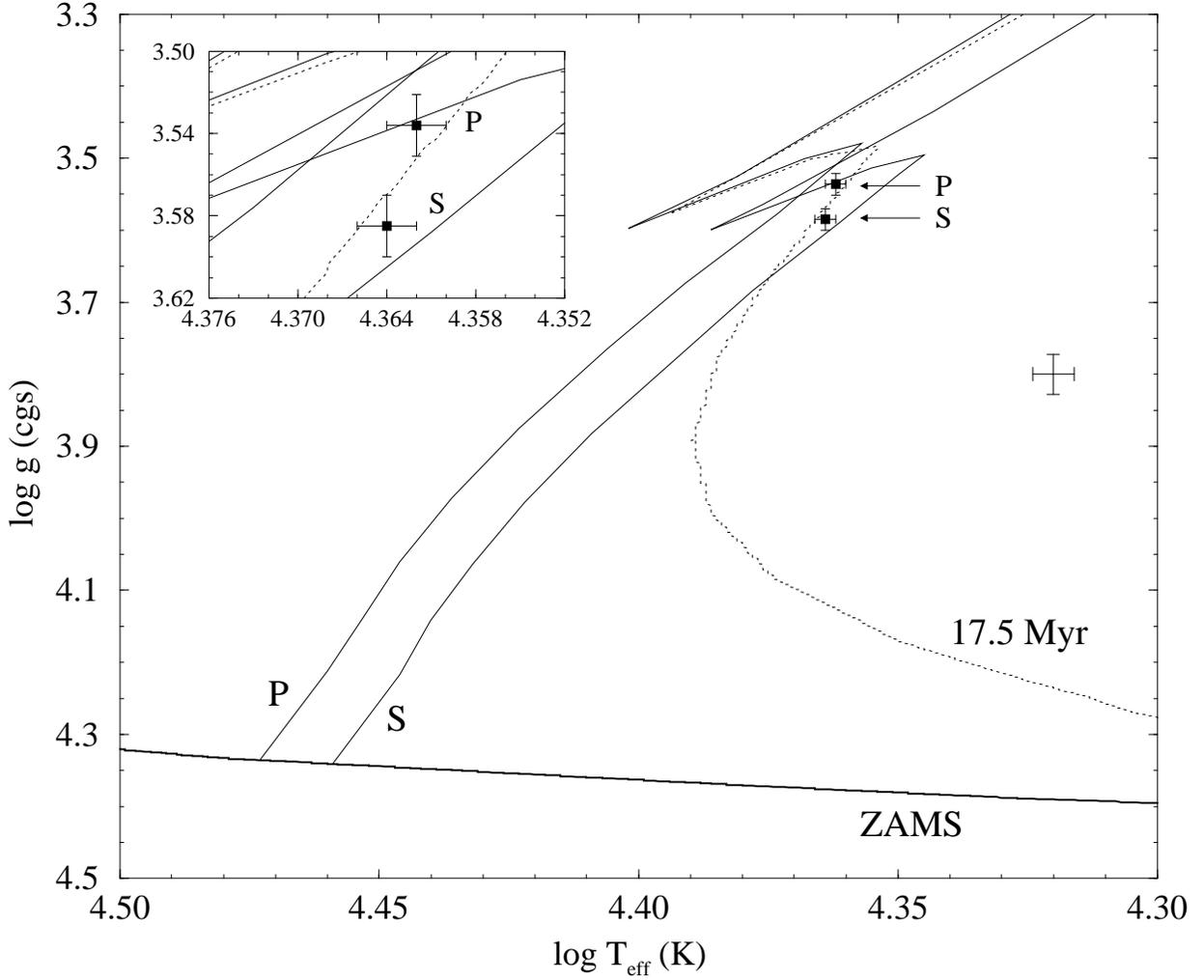}
\figcaption[fig8.eps]{$\log g - \log T_{\rm eff}$ plot of HV~2274 (P stands for
primary component and S for secondary component). The evolutionary tracks and 
isochrones were computed from the evolutionary models of CG that consider 
$\alpha_{\rm ov}=0.2$ and with $Z=0.007$ and $Y=0.26$. Mass loss has been 
taken into account when computing the evolutionary tracks. The error bars in 
the symbols only consider the ``differential'' effect in $\Delta \log g$ and 
$\Delta \log T_{\rm eff}$ (see text). The complete error budget is shown in 
the right part of the diagram. A blowup of the region where the stars are 
located is shown within the inset.
\label{fig:figEVtracks}}
\end{figure*}

The parameters yielded by the evolutionary models using the derived chemical
composition are listed in Table \ref{tab:modHV}. As observed, good agreement 
is obtained for the temperatures and the ages of both components. The 
evolutionary age of the system can be placed around 17.5~Myr. Also, notice 
that the stars have lost about 1\% of their initial mass via stellar winds. 
The location of both components of HV~2274 is shown in the $\log g-\log T_{\rm 
eff}$ plot (Fig. \ref{fig:figEVtracks}). Also shown in this figure are the 
evolutionary tracks at $Z=0.007$ and $Y=0.26$ for the ZAMS masses of the 
components and the isochrone that provides the best fit to the system. The 
agreement between the theoretical and observed stellar properties are well 
within the observational errors. It is also evident from the plot that the 
two stars are rather evolved and close to the terminal age main-sequence 
(TAMS). 

\placetable{tab:modHV}
\placefigure{fig:figEVtracks}

Unfortunately, HV~2274 is not a very suitable object for a detailed study of
the effects of overshooting through isochrone fitting, since both components
are very similar. This means that, as long as both stars lie within the main
sequence, it is always possible to find an initial helium abundance
that reproduces the observed properties of the components. There exists no
possibility of discriminating between different overshooting parameters since
their effect is small inside the main sequence and also it is very similar
for both components (small ``zero point'' changes but no ``differential
effect'').

\begin{figure*}
\epsscale{1.00}
\plotone{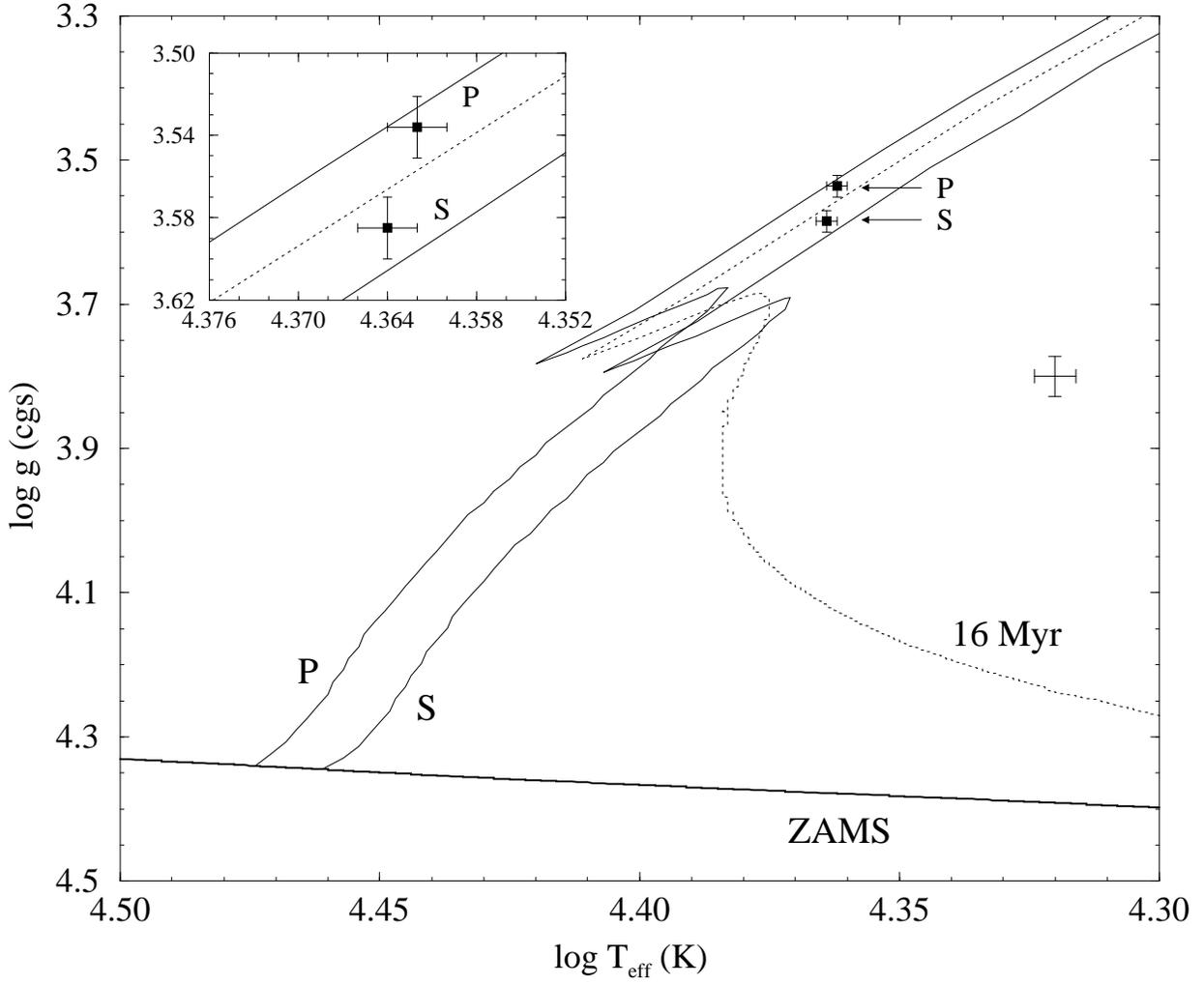}
\figcaption[fig9.eps]{Same as Figure \ref{fig:figEVtracks} but for 
the evolutionary models of CG that consider $\alpha_{\rm ov}=0$.
\label{fig:figEVtracks_st}}
\end{figure*}

However, in the case of $\alpha_{\rm ov}<0.2$, both components are located
beyond the TAMS. In such a situation, the evolution of the stars is so fast 
that, unless they have almost identical mass, their effective temperatures 
should be significantly different (at the same age). This is in disagreement 
with the results of the analysis of the light curves and the UV 
spectrophotometry.  Moreover, a mass ratio of unity is not compatible with 
the radial velocity curve solution. We show in Figure \ref{fig:figEVtracks_st} 
a $\log g - \log T_{\rm eff}$ of HV~2274 with the corresponding evolutionary 
tracks and best-fitting isochrone for $\alpha_{\rm ov}=0$ for $Y=0.26$ (as for 
$\alpha_{\rm ov}=0.2$, this value is the one that best reproduces the observed 
temperatures).

Stellar lifetime considerations can also be considered. Actually, 
straightforward calculations using evolutionary model tracks show that, 
for the masses of HV~2274 components, less than 1\% of the stellar lifetime 
is spent beyond the TAMS. This implies that it is very unlikely to observe 
both components of a binary system in the rapid evolving stage beyond the main
sequence. The argument constitutes additional evidence favoring 
evolutionary models with an overshooting parameter larger than 0.2, which 
predict both components to be located within the main sequence. For lower 
overshooting parameters, both components are predicted to lie in the 
Hertzsprung gap of the H-R diagram.

From all the previous evidences, an overshooting parameter below 0.2 can be
confidently ruled out. This result is especially valuable since it is the 
first test of the significance of overshooting that has been made for stars 
of such low metal abundance.

\placefigure{fig:figEVtracks_st}

A parallel test of stellar structure and evolution models can also be
performed, since HV~2274 has an eccentric orbit and W92 
have determined its apsidal motion rate. The apsidal motion rate of a binary 
system can be theoretically computed as the linear sum of a general 
relativity term (GR) and a classical term (CL). The latter, which is the 
most important contribution in close systems like HV~2274, depends on the 
internal mass distribution of the stars ($k_2$), which can also be 
theoretically estimated from the evolutionary models. Linear interpolation
in the CG models provides the values for $\log {k_2}$ shown in Table 
\ref{tab:modHV}. These values were corrected for stellar rotation by means of 
the formula of Claret \& Gim\'enez (1993). With this correction, $\Delta 
\log {k_2}_{\rm P,S}=-0.03$, the internal concentration parameters are 
$\log {k_2}_{\rm P}=-2.46\pm0.02$ and $\log {k_2}_{\rm S}=-2.43\pm0.02$, 
which lead to a mean system value of $\log {\overline k_2}=-2.45\pm0.02$. 
Also using the formulas in Claret \& Gim\'enez (1993), we derive a total 
theoretical apsidal motion rate of $\dot{\omega} 
{\rm (th)}=3.06\pm0.30$~deg/yr, in which ${\dot{\omega}}_{\rm CL}=
2.97$~deg/yr is the classical term and ${\dot{\omega}}_{\rm GR}=
0.09$~deg/yr is the general relativistic contribution. When comparing 
the theoretical determination with the observed value, $\dot{\omega} 
{\rm (obs)}=2.93\pm 0.07$~deg/yr, we find a remarkable agreement. The 
results of the apsidal motion study are summarized in Table 
\ref{tab:apsidHV2274}.

\placetable{tab:apsidHV2274}

The internal concentration of a star is strongly influenced by the convection 
parameters adopted for the stellar cores, in particular $\alpha_{\rm ov}$, but 
weakly dependent on the chemical composition (for a given set of 
convection parameters). In general, a larger overshooting means a higher 
internal concentration, since the star can remain a longer time on the main 
sequence. From the study of V380~Cyg (Guinan et al. 1999) and the paper by 
Claret \& Gim\'enez (1993) it can be deduced that, for the masses and 
evolutionary stage of HV~2274, an increasing of the overshooting parameter of 
$\Delta \alpha_{\rm ov}=0.1$ implies a decreasing of $\Delta \log k_2=-0.02$, 
with a good linear behavior. Taking into account this crude approximation and 
the $1\sigma$ uncertainties of the absolute dimensions, it is inferred that the 
observed apsidal motion rate is only reproduced if $0.1<\alpha_{\rm ov}<0.5$. 
Although the formal best fit is obtained for $\alpha_{\rm ov}=0.3$.

The overall conclusion that can be reached from the comparison of the 
observed properties with stellar evolutionary models is that the overshooting
parameter that best reproduces the observations can be constrained between 
$0.2<\alpha_{\rm ov}<0.5$. The constraint was possible by combining the 
analysis of the positions in the theoretical H-R diagram (lower limit) and the 
study of the apsidal motion rate (upper limit). We have to stress that this 
result is valid only for the component masses of HV~2274 and for the chemical 
composition of the system, $Z=0.007$ and $Y=0.26$. V380~Cyg (Guinan et al. 
1999) is a solar composition system ($Z=0.02$, $Y=0.28$) with a primary 
component of very similar mass to that of the components of HV~2274 (within 
10\%).  However, the overshooting parameter needed for reproducing observed 
properties of V380~Cyg is larger (and better constrained) than the value that 
we have found for HV~2274. Since a mass effect is ruled out, the only possible 
explanations are either an overshooting parameter variable (increasing) with 
evolution or a chemical composition effect. Both possibilities have 
interesting consequences and deserve further study.

\section{Tidal Evolution}

The observations show that the components of HV~2274 are moving in an 
eccentric orbit and that they appear to have rotational velocities synchronized 
with the orbital velocity at periastron. This is a common situation since in 
most cases the circularization timescale is much larger than the 
synchronization timescale. In order to compare this with the theoretical 
predictions, we adopted the tidal evolution formalism of Tassoul (1987, 1988). 
The mathematical expressions, which were taken from Claret et al. (1995), 
relate the time variation of the eccentricity and the angular rotation with 
the orbital and stellar physical properties: mass ratio, orbital period, mass, 
luminosity, radius and gyration radius. We stress the importance of 
integrating the differential equation that governs the variation with time of 
the eccentricity and the angular velocity rather than a simple timescale 
calculation (linear approximation). This is especially advisable for evolved 
systems like HV~2274, since the components undergo strong radius and 
luminosity changes. 

We therefore integrated the differential equations along evolutionary tracks
for the ZAMS masses of the HV~2274 components. The orbital period was assumed 
to be constant during the integration. As contour conditions, we imposed the 
eccentricity to be the observed value ($e=0.136\pm0.012$) at the present age 
of the system ($\tau=17.5\pm2.5$~Myr). Also, the angular rotation was 
normalized to unity at the ZAMS. The calculations were done with the 
evolutionary models at $Z=0.007$ and $Y=0.26$. As expected, the theoretical 
predictions indicate a system that has reached synchronism but 
that has not yet circularized its orbit. Indeed, both components reduced 
the difference between their angular velocity and the pseudo-synchronization 
angular velocity to less 0.1\% of the ZAMS value at the early age of 2~Myr. 
From the contour condition it is inferred that the ZAMS orbital eccentricity 
was quite large ($e_{\rm ZAMS} \approx 0.7$), and thus it has undergone a 
noticeable decreasing from this initially high value. The theory also predicts 
that the circularization will take place as soon as the primary component 
increases significantly its radius shortly after the onset of the shell 
hydrogen burning. The actual time, however, depends on the amount of convective 
overshooting, which affects the position of the TAMS.

\section{Conclusions}

This paper demonstrates clearly that the study of extragalactic eclipsing
binaries is now feasible and can lead to some interesting and important 
results on the properties and evolution for stars outside of our Galaxy.  
Also, as shown in our first paper, eclipsing binaries can serve as 
excellent ``standard candles'' for calibrating the distance to the LMC and 
in the near future, the Andromeda Galaxy. Remarkably, less than ten years 
ago there were fewer than 150 extragalactic eclipsing binaries known, with 
most of these being members of the LMC, SMC and M31.  However, this picture 
has drastically changed with the serendipitous discovery of thousands of new 
extragalactic eclipsing binaries from mainly the OGLE and MACHO microlensing 
surveys. For example, today there are nearly as many eclipsing binaries 
known in the LMC (2500) than cataloged in the Milky Way itself (about 
3000). In fact, it is estimated that a total of about 8000 eclipsing 
binaries will probably be discovered in the LMC from the MACHO program 
when it is completed (Cook 1999).
 
The large number of eclipsing binary systems provides a wide variety of
systems and stars to choose for further study. To get the maximum
scientific returns from these stars, however, requires considerable
effort, mainly because high quality light curves and radial velocity
curves are essential to extract the fundamental astrophysically
important information from these systems. Moreover, calibrated
spectroscopy or spectrophotometry (as was used here from HST), or
standardized $UBVRI$ or well-calibrated Str\"omgren $uvby$ photometry,
as well as high-dispersion spectroscopy are needed to determine the
temperatures and metalicities of the stars and also to determine
interstellar reddening and extinction needed for distances and
modeling. Although it takes a lot of work, none of the needed
observations is beyond current instrumental capabilities. Good-quality
light curves of 14$^{\rm th}$ and 15$^{\rm th}$~mag systems can
routinely be done with 1-m aperture telescopes. Spectroscopic
observations of these same stars (for radial velocity, temperature and
abundance studies), however,  need to be done with 4-m class telescopes
or, more efficiently, from space with HST using STIS.

In addition to HV~2274, we have secured HST/FOS spectra covering
$1150-4820$~\AA\/ for nine other LMC and SMC eclipsing systems with
components having late O and early B spectral types. The systems with
FOS spectra are HV~982, HV~1620, HV~1761, HV~1876, HV~2226, HV~2241,
HV~5936, HV~12634 and EROS~1044.  We also have IUE FUV spectra for a
several other LMC eclipsing binaries. Most of these systems have light
curves but lack satisfactory radial velocity curves. We have recently
submitted requests for time on the 4-m CTIO telescope and plan to
submit a proposal to the HST later this year for STIS observations.
 
In the near future we hope to revisit HV 2274 when better light and
radial velocity curves become available. Improved masses, stellar radii
and distances are possible with more observations. It may be somewhat
strange that the properties of this faint LMC system are better
determined than most eclipsing binaries in our own Galaxy. This
situation arises mainly because very few eclipsing binaries in our
Galaxies have FOS spectrophotometry from which accurate temperatures
can be derived.  Also, HV 2274 is an interesting target for
high-precision photometry since the component stars lie near the
locations in the H-R diagram in our Galaxy where $\beta$ Canis Majoris
variables are located. It would be interesting to conduct more
extensive photometry to see if light variations from pulsations of the
stars can be found. If short-term (several hours) light variations are
discovered, then HV 2274 could be a prime candidate for a detailed
asteroseismology study. In this case, the eclipses could be used to
isolate the star or stars that pulsate.
 
\acknowledgments

This work was supported by NASA grants NAG5-7113 and $HST$ GO-06683 and by 
NSF/RUI AST 93-15365. I. R. acknowledges the grant of the {\em Beques
predoctorals per a la formaci\'o de personal investigador} by the CIRIT
(Generalitat de Catalunya, Spain)(ref. FI-PG/95-1111). J. D. P. acknowledges 
the support of the New Zealand Marsden Fund for Curiosity-driven Research.

%
%

\clearpage

\begin{deluxetable}{cccc}
\tablewidth{0pt}
\tablecaption{HST/FOS spectra of HV~2274 in four different wavelength regions
obtained on 1996 November 15, when the eclipsing binary was near quadrature
($\phi=0\fp22$). In this situation, the combined flux of both components is
observed. \label{tab:logFOS}}
\tablehead{\colhead{Range (\AA)} & \colhead{res. (\AA/bin)} &
\colhead{$t_{\rm int}$ (sec)} & \colhead{S/N}}
\startdata
1140--1620   &   1.0  & 920  & 25 \nl
1550--2350   &   1.6  & 680  & 30 \nl
2220--3300   &   2.2  & 600  & 50 \nl
3210--4820   &   3.2  & 680  & 40 \nl
\enddata
\end{deluxetable}

\begin{deluxetable}{rrrrrr}
\tablewidth{0pt}
\tablecaption{HST/GHRS medium-resolution spectra of HV~2274. 
\label{tab:logGHRS}}
\tablehead{\colhead{Nr.} & 
\colhead{HJD-2450000} & \colhead{Phase\tablenotemark{a}} & 
\colhead{$\lambda$ (cent.)}}
\startdata
1 & 402.9159 & 0.2172 & 1335\nl
2 & 402.9733 & 0.2272 & 1300\nl
3 & 413.0330 & 0.9840 & 1335\nl
4 & 413.0689 & 0.9903 & 1300\nl
5 & 413.7106 & 0.1024 & 1335\nl
6 & 413.7589 & 0.1108 & 1300\nl
7 & 416.5881 & 0.6049 & 1335\nl
8 & 416.6324 & 0.6126 & 1300\nl
9 & 417.1551 & 0.7039 & 1335\nl
10& 417.8555 & 0.8263 & 1335\nl
11& 417.9199 & 0.8375 & 1300\nl
12& 420.2828 & 0.2501 & 1335\nl
13& 420.8290 & 0.3455 & 1335\nl
14& 420.8702 & 0.3527 & 1300\nl
\enddata
\tablenotetext{a}{The orbital phase is computed according to the linear 
ephemeris $\mbox{Min I} = 2448099.818 + 5\fd726006 \, E$. Heliocentric 
correction applied.}
\end{deluxetable}

\begin{deluxetable}{cccrccccr}
\tablewidth{0pt}
\tablecaption{Radial velocity measurements of HV~2274 from HST/GHRS 
medium-resolution spectra. The mean errors are about 15~km~s$^{-1}$. Wt. is 
the relative weight assigned to each measurement. O$-$C is the residual in the
radial velocity curve fit. \label{tab:RVHV2274}}
\tablehead{\multicolumn{4}{c}{P component} && 
\multicolumn{4}{c}{S component}}
\startdata
HJD$-$ &$v_{\rm r}$ &Wt.&\multicolumn{1}{c}{O$-$C} &&
HJD$-$ &$v_{\rm r}$ &Wt.&\multicolumn{1}{c}{O$-$C} \nl
2450000&\multicolumn{1}{c}{\footnotesize (km~s$^{-1}$)}& 
&\multicolumn{1}{c}{\footnotesize (km~s$^{-1}$)}&&
2450000&\multicolumn{1}{c}{\footnotesize (km~s$^{-1}$)}&
&\multicolumn{1}{c}{\footnotesize (km~s$^{-1}$)}\nl
\tableline
402.91589&170.3&0.5&   24.9&&402.91589&482.8&1.0& $-$7.0 \nl
402.97328&137.9&0.5& $-$7.8&&402.97328&505.4&1.0&   15.9 \nl
413.75888&210.2&1.0&   19.0&&413.71058&410.3&1.0&$-$20.1 \nl
417.15507&453.3&1.0&    4.8&&417.15507&154.4&1.0&$-$11.5 \nl
417.85554&466.9&1.0& $-$8.4&&417.85554&134.1&0.5& $-$3.9 \nl
417.91995&489.6&1.0&   16.4&&417.91995&143.8&1.0&    3.4 \nl
420.28279&149.9&0.5&    1.1&&420.28279&482.8&0.5& $-$3.6 \nl
420.82902&179.3&1.0&$-$10.4&&420.82902&419.4&1.0&$-$23.5 \nl
420.87019&182.0&1.0&$-$12.1&&420.87019&466.7&1.0&   28.4 \nl
\enddata
\end{deluxetable}

\begin{deluxetable}{lll}
\small
\tablewidth{0pt}
\tablecaption{Orbital and stellar properties of HV2274 derived from the 
analysis of the light and radial velocity curves and the FOS spectrum fit. 
The numbers in parentheses are the standard errors affecting the last 
digits of each parameter. Subscripts P and S refer to the primary and 
secondary components, respectively. \label{tab:HV2274}}
\tablehead{\multicolumn{3}{c}{Light curve solution}}
\startdata
$P=5.726006(12)$ days & $e=0.136(12)$ & 
$ \left. \frac{L_{\rm S}}{L_{\rm P}} \right|_{B} = 0.845(5)$\nl
$i=89\fdg6(1.3)$ & $\omega (1990.6)=73\fdg3(1.5)$ &
$ \left. \frac{L_{\rm S}}{L_{\rm P}} \right|_{V} = 0.844(5)$\nl
$r_{\rm P} = 0.255(2)$ & $\dot{\omega}=2.93(7)$~deg/yr &
$ \left. \frac{L_{\rm S}}{L_{\rm P}} \right|_{I} = 0.843(5)$\nl
$r_{\rm S} = 0.234(2)$& $\frac{{T_{\rm eff}}_{\rm S}}
{{T_{\rm eff}}_{\rm P}} = 1.005(5)$&\nl
\tableline
\tablevspace{1pt}
      \multicolumn{3}{c}{Radial velocity curve solution}\nl
\tableline
$K_{\rm P} = 166.2(5.9)$~km~s$^{-1}$ & $q=\frac{M_{\rm S}}{M_{\rm P}}=
0.938(45)$ & $\gamma = 312(4)$~km~s$^{-1}$ \nl
$K_{\rm S} = 177.3(5.8)$~km~s$^{-1}$ & $a=38.58(93)$~R$_{\sun}$ & \nl
\tableline
\tablevspace{1pt}
      \multicolumn{3}{c}{UV/Optical spectrophotometry} \nl
\tableline
${T_{\rm eff}}_{\rm P}=23000(180)$~K & $[Fe/H]=-0.45(6)$~dex
& $E(B-V) = 0.120(9)$\\
${T_{\rm eff}}_{\rm S}=23110(180)$~K & $\mu_{\rm PS}=1.9(7)$~km~s$^{-1}$ & 
\\
\tableline
\tablevspace{1pt}
      \multicolumn{3}{c}{Stellar physical properties}\nl
\tableline
\multicolumn{3}{c}{\(\begin{array}{lll}
M_{\rm P} = 12.2(7)~{\rm M}_{\odot}  &\mbox{\hspace{2cm}}& 
M_{\rm S} = 11.4(7)~{\rm M}_{\odot}  \\
R_{\rm P} = 9.86(24)~{\rm R}_{\odot} && R_{\rm S} = 9.03(24)~{\rm R}_{\odot} \\
\log g_{\rm P} = 3.536(27)~{\rm dex} && \log g_{\rm S} = 3.585(29)~{\rm dex} \\
{T_{\rm eff}}_{\rm P}=23000(180)~{\rm K} &
& {T_{\rm eff}}_{\rm S}=23110(180)~{\rm K}\\ 
\overline{\rho_{\rm P}}=0.018(2)~\mbox{g cm$^{-3}$} &
& \overline{\rho_{\rm S}}=0.022(2)~\mbox{g cm$^{-3}$}\\
{\rho_{\rm c}}_{\rm P}\tablenotemark{a}=4.2(6)~\mbox{g cm$^{-3}$} &
& {\rho_{\rm c}}_{\rm S}\tablenotemark{a}=4.8(7)~\mbox{g cm$^{-3}$}\\
{v_{\rm rot}}_{\rm P}=125~\mbox{km s$^{-1}$}&
&{v_{\rm rot}}_{\rm S}=115~\mbox{km~s$^{-1}$}
\end{array}\)
}\nl
\tableline
\tablevspace{1pt}
      \multicolumn{3}{c}{Distance determination\tablenotemark{b}} \nl
\tableline
\multicolumn{3}{c}{\(\begin{array}{lll}
d_{\rm HV2274} = 46.8(1.6)~{\rm kpc} && V_{\circ}-M_v = 18.35(7)~{\rm mag}\\
d_{\rm LMC} = 45.7(1.6)~{\rm kpc} && V_{\circ}-M_v = 18.30(7)~{\rm mag}
\end{array}\)
}\nl
\enddata
\tablenotetext{a}{Following Kopal (1959) and using the internal concentration 
parameters $k_2$ in Table \ref{tab:apsidHV2274}.}
\tablenotetext{b}{From Guinan et al. (1998b).}
\end{deluxetable}

\begin{deluxetable}{lrr}
\tablewidth{0pt}
\tablecaption{Results of a linear interpolation in the CG models that consider
$\alpha_{\rm ov}=0.2$. The input parameters are the observed absolute 
dimensions of HV~2274 and a chemical composition of $Z=0.007$ and $Y=0.26$. 
\label{tab:modHV}}
\tablehead{ &\multicolumn{1}{c}{P comp.}&\multicolumn{1}{c}{S comp.}}
\startdata
$T_{\rm eff}$ (K)    & $23260\pm680\hspace{3pt}$ & $22940\pm730\hspace{3pt}$ \nl
$Age$ (Myr)           & $16.7\pm1.6\hspace{6pt}$ & $18.3\pm1.9\hspace{6pt}$ \nl
$\log (L/L_{\odot})$  & $4.40\pm0.07$ & $4.30\pm0.08$ \nl
$\log k_2$            & $-2.43\pm0.02$& $-2.40\pm0.02$\nl
$M_{\rm ZAMS}$ (M$_{\sun}$)& $12.37\pm0.72$ & $11.53\pm0.72$ \nl
\enddata
\end{deluxetable}

\begin{deluxetable}{l}
\tablewidth{0pt}
\tablecaption{Summary of results of the apsidal motion study of HV~2274.
Evolutionary models of CG with $\alpha_{\rm ov}=0.2$, $Z=0.007$ and $Y=0.26$ 
were used for the theoretical calculations. The superscript r means 
that the rotational correction has been applied ($\Delta \log {k_2}_{\rm P,S} 
({\rm rot.})= -0.03$).\label{tab:apsidHV2274}}
\tablehead{\colhead{Apsidal motion}}
\startdata
\( \left. \begin{array}{l}
\log {k_2}^{\rm r}_{\rm P} ({\rm th})= -2.46(2) \\
\log {k_2}^{\rm r}_{\rm S} ({\rm th})= -2.43(2)
\end{array} \right\}
\Longrightarrow \log {{\overline{k}}_2}^{\rm r} ({\rm th}) = -2.45(2)\) \nl
\( \left. \begin{array}{l}
{\dot{\omega}}_{\rm CL} =2.97(30) \;{\rm deg/yr} \\
{\dot{\omega}}_{\rm GR} =0.09(0) \;{\rm deg/yr}
\end{array} \right\}
\Longrightarrow \dot{\omega} ({\rm th})=3.06(30) \;{\rm deg/yr} \)\nl
\hspace{5.43cm}$\dot{\omega} ({\rm obs})=2.93(7) \;{\rm deg/yr}$\nl
\enddata
\end{deluxetable}

\begin{figure*}
\epsscale{1.00}
\plotone{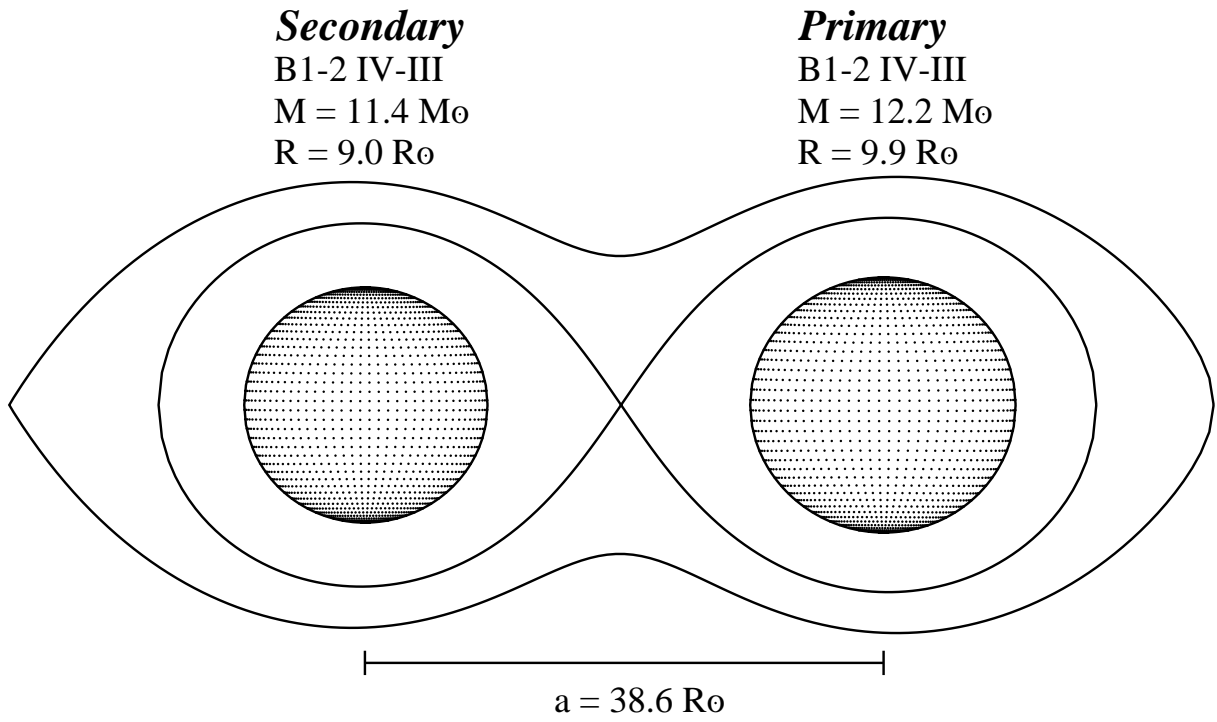}
\figcaption[fig7.eps]{3-D picture of HV~2274 in quadrature phase.
\label{fig:fig3D}}
\end{figure*}

\end{document}